\documentclass[preprint]{revtex4}
\usepackage{graphics}
\usepackage{amssymb}

\newcommand{\be}{\begin{equation}}
\newcommand{\ee}{\end{equation}}
\newcommand{\ba}{\begin{array}}
\newcommand{\ea}{\end{array}}
\newcommand{\bea}{\begin{eqnarray}}
\newcommand{\eea}{\end{eqnarray}}
\def\bra{\langle}
\def\ket{\rangle}

\begin{document}

%
%
\title{TRANSIENT PATTERN FORMATION IN A STOCHASTIC MODEL OF CANCER GROWTH}

\author{Anna \surname{Ochab--Marcinek}}
\affiliation{Marian~Smoluchowski Institute of Physics,
 Jagellonian University, Reymonta~4, 30--059~Krak\'ow, Poland}
\email{ochab@th.if.uj.edu.pl}

\date{13 May 2005}

\begin{abstract}
We study a spatially inhomogeneous model of cancer growth based on Michaelis--Menten kinetics,
subjected to additive Gaussian noise and multiplicative dichotomous noise.
In presence of the latter, we can observe a transition between two stationary states of the system,
The transient behaviour generates a spatial pattern of two phases, where cancer cells or immune
 cells predominate.
\end{abstract}
\maketitle

\section{Introduction}

The presence of noise in biological systems may be regarded not only as a mere source of disorder
 but also as a factor which introduces positive and organising rather than disruptive changes
 in the system's dynamics. Some of the more important
 examples of noise induced effects are: stochastic resonance \cite{Gammaitoni_Hanggi},
 resonant activation \cite{Doering_Gadoua,Iwaniszewski,Ochab_Gudowska},
 noise enhanced stability  \cite{Spagnolo},
 stochastic hysteresis and synchronisation \cite{Mahato, Gudowska}, bifurcation effects \cite{Klosek}
 or pattern formation \cite{Valenti_Fiasconaro_Spagnolo, Ochab}.\\

  The effect of cell-mediated immune surveillance against
 cancer \cite{Garay, Stepanova, Vladar} may be a specific illustration of the coupling
  between noise and a biological system.
  The series of reactions between cytotoxic cells and a tumour tissue may be
  considered to be well approximated \cite{Garay} by a saturating, enzymatic-like process
  whose time evolution equations are similar to the standard Michaelis--Menten kinetics.
  Random variability of kinetic parameters defining that process may affect
the extinction of the tumour \cite{Garay,Prigogine_Lefever}. In the present paper,
we investigate the  effect of pattern formation in a spatially inhomogeneous model of cancer
growth. We describe the dynamics of the system in terms of chemical
 reactions.\\

Biochemical reactions are usually described in terms of
phenomenological kinetic rates formulated by standard
stoichiometric analysis. In such deterministic models, molecular
fluctuations can be incorporated by including additional source of
stochastic fluxes  represented e.g. by the additive white noise:

\be \label{eq:additive}
 \frac{dx}{dt}=f(x)-g(x)+\xi(t)\ . \ee

The above Langevin
equation is based on a continuous description of molecular species:
time evolution of an input or output concentrations $x$ produced
at a rate $f(x)$ and degraded at rate $g(x)$ defines a
deterministic flux of reacting species and can be used when
modelling processes involve sufficient concentrations of reacting
agents.\\

 When, in turn, only a few molecules are present, discrete
structure of single-reaction events becomes important and the
evolution of the system is better described by an appropriate
Master equation \bea
 P_{i}(t+\Delta
t)=[f_{i-1}\Delta t + {\cal{O}}(\Delta t)]P_{i-1}(t)
+ [1-(f_i+g_i)\Delta t + {\cal{O}}(\Delta t)]P_{i}(t) \nonumber \\
+[g_{i+1}\Delta t+ {\mathcal{O}}(\Delta t)]P_{i+1}(t),\nonumber
\label{ME}
 \eea where $P_i(t)$ is the probability for there being $i$ molecules
(objects) at time $t$, $f_i$ is a probability per unit time
of transition $i\rightarrow i+1$ and $g_i$ is a probability per unit time
of the opposite transition $i\rightarrow i-1$. Using (\ref{ME}) one can
formulate the equation for the first moment $\bra i \ket=\sum i P_{i}(t)$
and then apply a limiting procedure for converting it
 into a partial differential equation. The procedure
involves introducing a small parameter $\epsilon$ and letting
$x=\epsilon i$ and $P_{i}\equiv P(x,t|y,0)$.
Here $P(x,t|y,0)$ is the probability that the random variable has
the value $x$ at time $t$ given it had the value $y$ at $t=0$. By
considering now a sequence of birth-and-death processes such that
$ \epsilon (f_i(\epsilon)-g_i(\epsilon))=a(i\epsilon) +
\mathcal{O}(\epsilon)$ and $\epsilon^2(f_i(\epsilon)+g_i(\epsilon))=b(i\epsilon) +
\mathcal{O}(\epsilon)$.
Letting $\epsilon\rightarrow 0$ and using the above conditions, we arrive at the
Fokker--Planck equation
\bea
\frac{\partial}{\partial t} P(x,t|y,0)=-\frac{\partial}{\partial
x}[a(x)P(x,t|y,0)]+\frac{\partial^2}{\partial x^2}[b(x)P(x,t|y,0)]\ .
\label{fokker}\eea
 We are usually
 interested in the effects of stochasticity for $i$ not too small.
Therefore, according to the above scheme, with increasing number
of reactants, stochastic effects analysed via the ME (or FPE) are
asymptotically equivalent to those described by the Langevin
equation with a multiplicative noise term \cite{kurtz}. \bea
 \frac{dx}{dt}
=f(x)-g(x)+\sqrt{f(x)+g(x)}\xi_t =a(x)+ \sqrt{b(x)}\xi_t
 \eea From amongst the presented possible ways of description of chemical and biological processes,
 we have chosen an approach using a chemical Langevin equation
with an additive driving noise term as in Eq. (\ref{eq:additive}) to demonstrate
the positive role of both additive and multiplicative noises in a
 regulatory model of the catalytic reaction. The description of cancer growth kinetics
is based on the phenomenological Michaelis--Menten scheme for
 the catalysis accompanying a spontaneous replication of cancer cells.\\


\section{The Model}
The interaction between cancer cells and cytotoxic cells
will be described by use of the predator-prey model
based upon the Michaelis--Menten kinetic scheme \cite{Garay_Lefever,Lefever_Horsthemke,Horsthemke_Lefever,
Prigogine_Lefever,Mombach,Ochab_Gudowska}. This model is a classical one and has been
extensively studied since the 1970s. Its validity has been verified experimentally e.g. in
\cite{Garay_Lefever}, where the authors examined the mechanism of immune rejection of a tumour
induced by Moloney murine sarcoma virus. The behavior of the cellular populations may be represented by the following scheme:
\be
X\ +\ Y\ \longrightarrow \!\!\!\!\!\!\!\! ^{k_{1}} \ \ \ \ Z\ \longrightarrow \!\!\!\!\!\!\!\! ^{k_{2}}\ \ \ \  Y\ +\ P\ \longrightarrow \!\!\!\!\!\!\!\! ^{k_{3}}\ \ \ \  Y.
\ee
$X$ represents here the population of tumour cells. $Y$, $Z$ and
$P$ are populations of active cytotoxic cells, bound cells and dead tumour cells,
 respectively. Cytotoxic cells bind the tumour cells at rate $k_{1}$;
subsequently, the cancer cells which have been bound are
killed and the complex dissociates at rate $k_{2}$; finally, dead cancer cells
decay at a rate $k_3$. The mechanism of cell replication will be first presented in one-dimensional case:
Cancer cells
replicate at a rate $\lambda$, but the process is being inhibited due to a limited volume of a
compartment, whose maximal carrying capacity is denoted by $N$.
We divide a one-dimensional space into small compartments of size $\Delta r=Na$, where $a$ is a
typical cell diameter. The overflow of replicated cells
may diffuse to neighbouring compartments at a rate proportional to the volume available there.
 If $X_i$ is the number of cancer cells occupying the $i$th compartment at time $t$, then, after a discrete time step $\Delta t$:

\be \label{discrete}
\begin{array}{lll}
X_i(t+\Delta t) & = & X_i + \lambda \Delta t X_i \left( \frac{N-X_i-P_i}{N}\right)\\
&&\\
 &  & + \lambda \Delta t \left( X_{i-1} \frac{N-X_i-P_i}{N} +X_{i+1} \frac{N-X_i-P_i}{N} \right.\\
 &&\\
 &  & \left. - X_i \frac{N-X_{i-1}-P_{i-1}}{N} - X_i \frac{N-X_{i+1}-P_{i+1}}{N} \right).
\end{array}
\ee

An analogous formula can be written for a two-dimensional case. The conversion of such an equation
into a continuous form will yield a logistic term (saturating growth) and diffusion terms.
Additionally, we assume that immune cells can diffuse at a rate proportional to a coefficient $D$. External
 environmental fluctuations will be modelled by the additive Gaussian noise $\xi(\overrightarrow r,t)$
  with autocorrelation
$<\xi(\overrightarrow r,t)\xi(\overrightarrow r',t')> = \delta(t-t')\delta(\overrightarrow r-\overrightarrow r')$.
We assume that since $\xi(\overrightarrow r,t)$ is an environmental noise (modelling e.g.
local temperature changes),
 it acts in the same way on all system variables. Fluctuations in
immune response will be introduced as a multiplicative dichotomous Markovian noise
$\eta(t)=\pm\Delta$ with mean frequency
 $\gamma$ and autocorrelation $<\eta(t)\eta(t')> = \frac{\Delta^2}{2}e^{-2\gamma|t-t'|}$. The spatio-temporal evolution of the tumour due to the above processes will be then
described by a set of balance equations with stochastic components:

\be\label{sim_system}
\left\{
\begin{array}{lll}
\frac{\partial x}{\partial t}&=&\lambda[1-(x+p)]x-(k_{1}+\eta(t))yx+\\
 & & +\lambda(1-p)(Na)^2\nabla^2 x +\lambda x(Na)^2\nabla^2 p + \sigma\xi(\overrightarrow r,t)\\
\frac{\partial y}{\partial t}&=&-(k_{1}+\eta(t))yx+k_{2}z+D\nabla^2 y  + \sigma\xi(\overrightarrow r,t)\\
\frac{\partial z}{\partial t}&=&(k_{1}+\eta(t))yx-k_{2}z  + \sigma\xi(\overrightarrow r,t)\\
\frac{\partial p}{\partial t}&=&k_{2}z-k_{3}p + \sigma\xi(\overrightarrow r,t)\ .
\end{array}
\right.
\ee

The $x(\overrightarrow r, t) ,y(\overrightarrow r, t),z(\overrightarrow r, t)$ and $p(\overrightarrow r, t)$
are local densities of cells at point $\overrightarrow r$. We assume that
the noise intensity $\sigma$ is the same for each variable of the system.
Both noises are assumed to be statistically independent: $<\xi(\overrightarrow r,t)\eta(s)>=0$.


\section{Stability Analysis}\label{sec:stability_analysis}
The stationary points $\{ x^*,y^*,z^*,p^*\}$ of the system are given by:
\be\label{stationary_points_1}
\{0,y,0,0\}
\ee
and
\be\label{stationary_points_2}
\left\{ \frac{k_3}{\lambda}\frac{\lambda-k_1 y}{k_3+k_1 y}, y,
\frac{k_1 k_3 y}{\lambda k_2}\frac{\lambda-k_1 y}{k_3+k_1 y},
\frac{k_1 y}{\lambda}\frac{\lambda-k_1 y}{k_3+k_1 y} \right \},
\ee
where the value of $y$ is defined by the condition $y(t)+z(t)=const.=E$. The sets of stationary points form two branches in the
 $x$-$y$-$z$-$p$-space.
 The branch (\ref{stationary_points_1}) changes its stability at the point
  $\{0,\frac{\lambda}{k_1},0,0\}$. It is repelling for
  $0 < y < \frac{\lambda}{k_1}$
  and attracting
  for $\frac{\lambda}{k_1} < y < 1$.
  The branch (\ref{stationary_points_2}) is attracting for
  $0 < y < \frac{k_3}{k_1}\left(-1+\sqrt{1+\frac{\lambda}{k_1}}\right)$.
  For $ \frac{k_3}{k_1}\left(-1+\sqrt{1+\frac{\lambda}{k_1}}\right) < y < 1$
   it consists of saddle points \cite{Ochab}. When $E,k_2,k_3$ are fixed,
    stability properties of the system depend
on the $k_1$ parameter (connected with the immune response efficiency) which is
  controlled by the dichotomous noise. As the value $k_1+\eta(t)$ changes, the branch
  (\ref{stationary_points_2}) moves up and down \cite{Ochab}.

\section{Simulation}\label{sec:simulation}

We have solved the stochastic differential equations (\ref{sim_system}) numerically,
 using their discrete form analogous to (\ref{discrete}) (i.e. the Euler scheme), on a $128 \times 128$ square lattice with periodic boundary conditions
 for $\overrightarrow r$. According to the statistical properties of $\eta(t)$,
  the waiting time between two switchings was generated from the exponential distribution. Since
   $x,y,z$ and $p$ are densities, their values never can be greater than $1$ or less than $0$.
Consequently, we applied reflecting boundaries at those values.\\

We performed a simulation with the following values of parameters:
\be
\begin{array}{l}
\lambda = 0.5,\  Na=1,\ D = 0.05,\  \sigma = 0.01,\  \Delta = 0.5,\\
 k_1 =1.75,\  k_2 = 0.1,\  k_3 = 0.1,\  \gamma = 0.01.
\end{array}
\ee \label{eq:initial}
$\gamma$ is the mean rate of switching in $\eta(t)$.
Simulation time: $T=20000$.
Initial conditions:
\be \label{eq:initial_point}
x(\overrightarrow r,0)=0,\  y(\overrightarrow r,0)=0.4,
\  z(\overrightarrow r,0)=0,\  p(\overrightarrow r,0)=0.
\ee

The values of parameters and initial conditions have been chosen so that we
could obtain a distinct spatial pattern:
The immune response rates $k_1+\Delta$ and $k_1-\Delta$, along
with $\lambda$ lead to two different types of stationary behaviour (see Sec. \ref{sec:results}).
The mean switching rate
is one or two orders of magnitude slower than other kinetic parameters, which gives the system a
possibility to approach the stationary states. The environmental noise intensity $\sigma$ has been
chosen in such a way that, combined with $\eta(t)$, it allows the system to jump between both mentioned states. The noise is,
however, weak enough to let the system form a pattern (otherwise the Gaussian noise would
dominate the picture) \cite{spagnolo_appb}. At the selected value of $D$, the pattern
has sufficiently distinct boundaries and is relatively stable (at higher values of $D$ it
would dissolve quickly, whereas at smaller $D$ it would form small ``grains"). Values of parameters $k_2$ and $k_3$ have been chosen by
 a trial-and-error procedure: $k_2$ is responsible for
the dissociation of $z$ into $y+p$. If the dissociation rate is large, then the active immune cells $y$ are
being released faster and thus the immune response is more effective. The $k_3$ parameter determines the
rate at which dead cancer cells are eliminated. Since dead cells occupy the living space,
this parameter controls the effective replication rate of
cancer cells.

\section{Results}\label{sec:results}

\begin{figure}[t]
\centering{
	\resizebox{10cm}{!}{\includegraphics{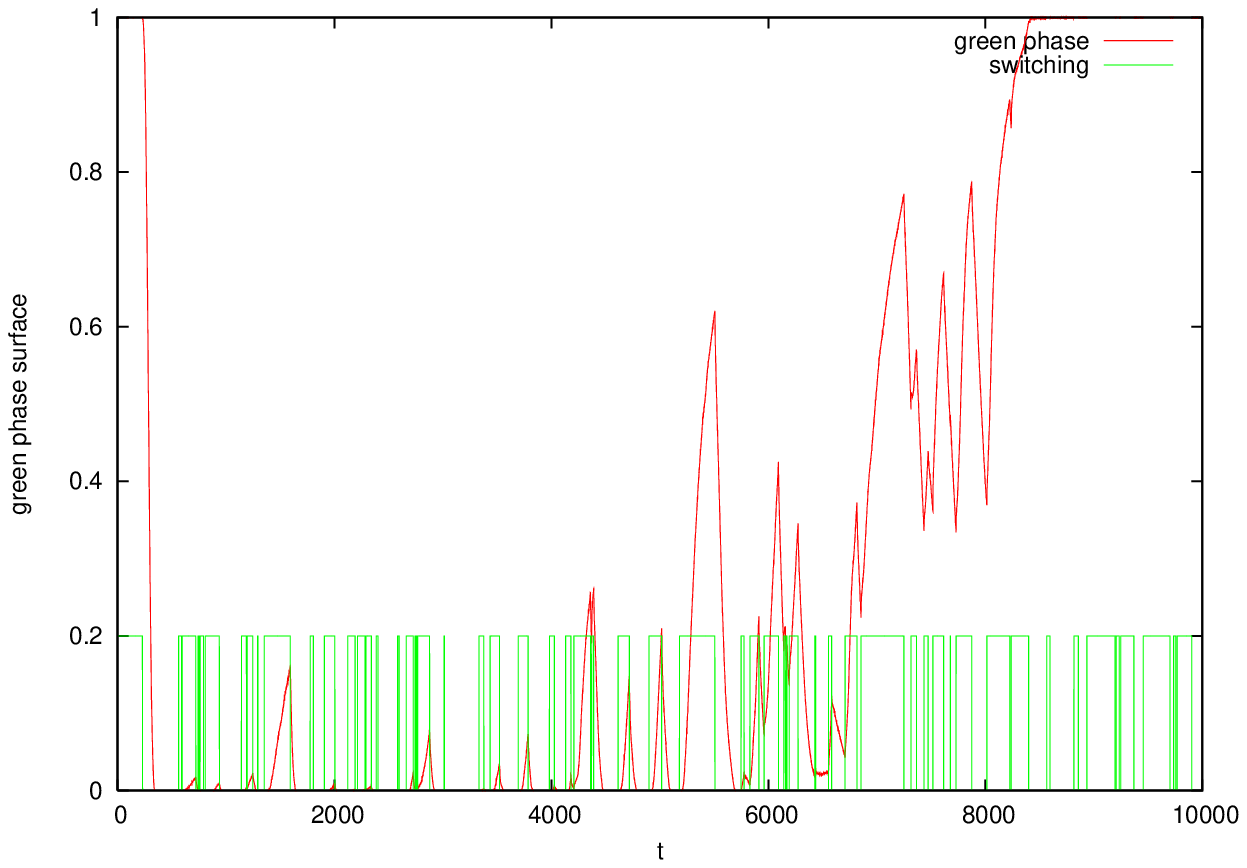}}\\
	\resizebox{10cm}{!}{\includegraphics{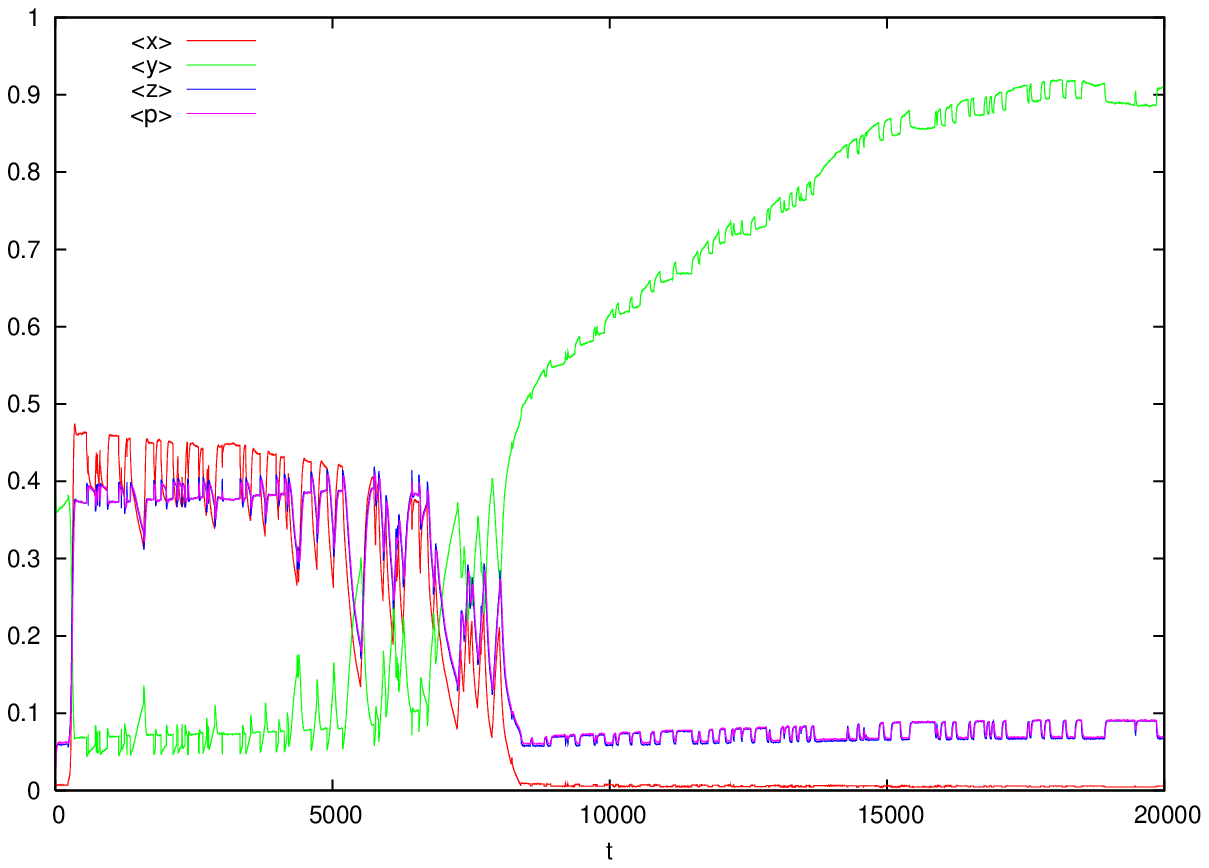}}
	}
\caption{Left: Total area of the ``$y$-phase" (counted points on the lattice where $x$
 is smaller than 0.03).
Right: Average densities of $x$, $y$ and $z$ populations over the simulation area.}
\label{fig:gps_average}
\end{figure}

The system driven by the additive Gaussian noise in the absence of the dichotomous switching
($\eta(t) = const = +\Delta$ or $\eta(t) = const = -\Delta$) fluctuates in the neighbourhood
of one of the two attracting branches,
  without jumping between them \cite{Ochab}. When the dichotomous noise is present,
  a coexistence of the two "phases"
is possible. Some parts of the system (the ``$x$-phase") can fluctuate around the
lower branch, whereas other parts (the ``$y$-phase") fluctuate in the vicinity of
the upper one. We observe the emergence of clusters, where immune cells $y$,
or, respectively, cancer cells $x$ predominate. The phase boundaries move back and forth
 depending on
 the dichotomous changes in the immune response intensity \cite{Ochab}. We observe several
 episodes of sudden formation and decay of "$y$-phase" clusters (Fig. \ref{fig:gps_average})
  until the "$y$-phase" prevails globally (Fig. \ref{fig:gps_average}).


\section{Conclusions}\label{sec:conclusions}

The measurement of the ``$y$-phase" total area and the average population densities over
the simulation area (Fig.\ref{fig:gps_average}) allow to notice that the spatially extended
system, as a whole, posseses two stationary states.
We can observe the transition between them:
 Shortly after starting from $x=0,\ y=0.4,\ z=0,\ p=0$, the system falls down onto the ``$x$-phase" branch.
The average population densities fluctuate around a certain constant level
(the first stationary state).
After some time, clusters of ``$y$-phase" begin to emerge and disappear, which is caused by
 the dichotomous changes in the position of the stationary branch (\ref{stationary_points_2}).
The average population level of $x$, $y$, $z$ and $p$ fluctuates strongly. As time goes on,
clusters form more and more frequently whereas their area becomes larger
and larger. Finally, the whole system escapes onto the ``$y$-phase" branch and climbs up \cite{Ochab}
towards
a new stationary state (fluctuations between reflecting boundaries of $x=0,\ y=1,\ z=0$), in which
the average population densities approach a new constant level, about which they weakly fluctuate
 (the second stationary state).


\end{document}